\documentclass[10pt,twocolumn,prl]{revtex4-1}
\usepackage{graphicx}
\usepackage{amssymb}
\usepackage{amsmath}

\begin{document}

\title{Transmission and diffraction properties of a narrow slit in ideal metal}

\author{B.\ Sturman$^1$, E.\ Podivilov$^1$, and M.\ Gorkunov$^2$}
\affiliation{$^1$Institute of Automation and Electrometry,
Russian Academy of Sciences, 630090 Novosibirsk, Russia \\
$^2$Shubnikov Institute of Crystallography, Russian Academy of
Sciences, 119333 Moscow, Russia}

\begin{abstract}

By solving Maxwell equations with the ideal-metal boundary conditions in the TM case, we
have fully described the transmission and diffraction properties of a single slit
regardless of its width. Efficiencies of the main transformation processes --
transmission, diffraction, and reflection -- are analyzed in the sub-to-few-wavelength range showing a number of sharp fundamental features. Close links with the case of real metal are considered.

\vspace*{1mm}


\end{abstract}

\maketitle

Solutions to electromagnetic diffraction problems based on the ideal-metal boundary conditions date back to the works of Sommerfeld, Rayleigh, and Bethe~\cite{Rayleigh,Sommerfeld,Bethe}, see also~\cite{Bouwkamp,Landau}. Such solutions are few in number, complicated, but indispensable in the subwavelength/nano optics where the usual Huygens-principle-based theory and intuitive approaches fail. Importantly, they link together the far- and near-field properties of the
electromagnetic field including the corner singularities~\cite{Meixner}.

The general upsurge of interest in nanooptics gave rise to many research areas involving nanostructured metals, such as near-field microscopy~\cite{Betzig}, metal-clad cavities~\cite{Hill}, bio-sensing~\cite{Anker}, and extraordinary light transmission (ELT) through nano-holes~\cite{Ebbesen98,Nature07}. Already the formulation of the corresponding problems strongly differs from that typical of classical optics: Instead of diffraction from plain obstacles~\cite{Bouwkamp}, one deals with funneling of light into (out of) apertures, with coupling of the opposite metal interfaces, with a local enhancement of the electromagnetic fields. Involvement of the surface plasmons in real metals further enriches the physics~\cite{Nature03}.

Theoretical basis of nanooptics of metals is a big issue. On the one hand, direct numerical methods typically map a tiny part of the actual space of variable parameters of the system with no real insight into the physics of multiscale phenomena. On the other hand, they are capable of a dramatic enhancement of the analytical tools providing virtually exact solutions to key physical problems. Reduction of complex problems to the basic elementary ones is a strong line of the studies, as it is known, e.g., for the ELT case~\cite{Garcia02,LalanneNature08,We3}: The transmittance of a perforated opaque metal film can often be accurately expressed by the efficiencies of the elementary single-interface transformation processes, while the film thickness trivially affects the positions of the Fabry-Perot resonances.

In this letter, we present a full-scale solution to the single-slit problem within the paradigm of ideal metal. This problem is among the most basic ones in nanooptics of metals. Our solution reveals a wealth of subwavelength and near-subwavelength features which have never been known. While our "ideal" problem is easier than that for real metals (owing, e.g., to the absence of the surface plasmons), it is more complicated than the Sommerfeld problem of diffraction from a single metal wedge~\cite{Sommerfeld}. Furthermore, it is applicable to almost real metals.

An important property of the slit geometry is the survival of a single fundamental propagating mode in the subwavelength case for ideal and real metals. This property is inherent in any multiply connected aperture cross-section~\cite{Landau,Jackson}, such as, e.g., a circular slit. Our results on the transmission properties are thus generic for a wide class of geometries.

\begin{figure}[h]
\centering
\includegraphics[width=8.3cm,height=2.4cm]{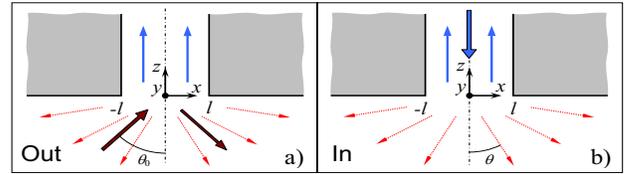}
\vspace*{-2mm}
\caption{ Two basic interface problems, "Out" and "In", corresponding to the incidence from outside (a) and inside (b).}\label{Fig1}
\end{figure}

Two distinct cases, "Out" and "In", for a slit of the half-width $l$ are depicted in Figs.~1a and 1b. In the case "Out", a plane wave of a unit amplitude at the wavelength $\lambda$, incident at the angle $\theta_0$, excites propagating modes in the slit, diffracted waves in air, and a reflected wave. In the case "In", a unit-amplitude propagating mode traveling to the interface $z = 0$ reflects back and excites diffracted waves in air.

In the TM case, the magnetic field amplitude has only a $y$-component $H = H(x,z)$, and the nonzero components of the electric field are $E_x = (-i/k_0)\,\partial H/\partial z$ and $E_z = (i/k_0)\, \partial H/\partial x$, where $k_0 = 2\pi/\lambda$. At the air-metal boundaries, the tangential component of $\vec{E}$ and the normal component of $\vec{H}$ turn to zero.

As a full set of eigenfunctions for $z > 0$ we choose~\cite{Landau}
\begin{equation}\label{h}
h_{\nu}(x) = \Big\{
\begin{array}{l}
\hspace*{-0.5mm}\cos (\pi\nu x/2l)\, ,  \hspace*{12mm} (\nu = 0, 2, \ldots)  \\
\vspace*{-3.5mm} \\
\hspace*{-0.5mm}\sin (\pi\nu x/2l) \;, \hspace*{12mm} (\nu = 1, 3, \ldots) \,. \nonumber
\end{array}
\end{equation}
\noindent The corresponding eigenvalues are $\beta_{\nu} = \sqrt{k_0^2 -(\pi\nu/2l)^2}$. The real and imaginary $\beta_{\nu}$ refer to the propagating and evanescent modes. For the fundamental propagating mode, $\nu = 0$, we have $\beta_0 = k_0$ and $h_0(x) = 1$. For $l < \lambda/4$ all other modes are evanescent.

For the case "Out" and $z \gtrless 0$, the field $H(x,z)$ is generally presented by the expansions
\begin{eqnarray}\label{Expansions}
H^> &=& \sum\limits_{\nu}
\;c_{\nu}\,b_{\nu} \;h_{\nu}(x)\;e^{\displaystyle i \beta_{\nu}z} \;,  \\
H^< &=& 2e^{\displaystyle ik_{0x}x}\cos(k_{0z}z) + \hspace*{-1.5mm}
\int_{-\infty}^{\infty} \hspace*{-1mm} a_k \, e^{\displaystyle ikx - i\varkappa_kz} \,
dk , \nonumber
\end{eqnarray}
\noindent where $k_{0x} = k_0\sin\theta_0$, $k_{0z} = k_0\cos\theta_0$, $\varkappa_k =
(k_0^2 - k^2)^{1/2}$, $c_0 = 1$, and $c_{\nu} = 2$ for $\nu \neq 0$. For $z > 0$ and $|x| \leq l$, we have satisfied Maxwell equations and the boundary conditions at $|x| = l$. Furthermore, we have $b_{\nu} = \langle H^>(x,0)\,h_{\nu} \rangle$, where $\langle .. \rangle$ indicates averaging over the slit. In the air region, $z < 0$, we have satisfied so far only Maxwell equations. Here the waves with $k < k_0$ are propagating, while for $k > k_0$ they are evanescent.

The amplitudes $b_{\nu}$ and $a_k$ can be found if we satisfy the remaining boundary conditions: $E_x^<(x,0) = E_x^>(x,0)\,\Theta(l -|x|)$, where $\Theta(x)$ is the Heaviside step function, and $H^<(x,0) = H^>(x,0)$ for $|x| < l$. They are equivalent to the relations
\begin{eqnarray}\label{akbnu}
a_k &=& - \frac{l}{2\pi\varkappa_k}\;\sum_{\nu} \; c_{\nu}\, \beta_{\nu}\,
b_{\nu}\,f_{\nu,k} \\
b_{\nu} &=& f_{\nu,\, -k_{0x}} + \frac{1}{2} \int_{-\infty}^{\infty} a_k\,f_{\nu,\,-k}
\;dk \;, \nonumber
\end{eqnarray}
\noindent where $f_{\nu,k} = \mbox{sinc}(kl + \pi\nu/2) + \mbox{sinc}(kl - \pi\nu/2)$ for $\nu = 0, 2, \ldots$, $f_{\nu,k} = i\,\mbox{sinc}(kl + \pi\nu/2) - i\,\mbox{sinc}(kl - \pi\nu/2)$ for $\nu = 1, 3, \ldots$, and $\mbox{sinc} (..) \equiv \sin (..)/(..)$. For even/odd $\nu$, $f_{\nu,k}$ is real/imaginary and even/odd in~$k$. Combining Eqs.~(\ref{akbnu}), we come to the set of coupled-mode equations
\begin{equation}\label{MasterObl}
b_{\nu} + \sum_{\nu'}\; T_{\nu \nu'}\,b_{\nu'} = f_{\nu, -k_{0x}}
\end{equation}
\noindent with the coupling coefficients
\begin{equation}\label{MatrixObl}
T_{\nu \nu'} = \frac{l}{4\pi} \,\beta_{\nu'} c_{\nu'} \,\int_{-\infty}^{\infty}
\,\frac{f_{\nu,-k}\,f_{\nu',k}}{\varkappa_k} \;dk \;.
\end{equation}
\noindent Obviously, $T_{\nu \nu'} = 0$ for the modes of different parity, i.e., the set~(\ref{MasterObl}) splits into two sets -- for the even and odd modes. Calculating $b_{\nu}$ from Eq.~(\ref{MasterObl}) and using Eq.~(\ref{akbnu}) for $a_k$, we solve {\it completely} the problem "Out". At $\theta_0 = 0$, the amplitudes $b_{\nu}$ are nonzero only for the even modes and the driving force is $f_{\nu,0} = 2\delta_{\nu 0}$.

Using Eqs.~(\ref{Expansions},\,\ref{akbnu}) and the integral representations~\cite{Bateman} of the Hankel function $H^{(1)}_0$, one can prove lastly the general relation for the diffracted component of $H^<$:
\begin{equation}\label{deltaH}
\hspace*{-2mm} H^<_d = -\frac{k_0}{2} \hspace*{-1mm} \int_{-l}^l \hspace*{-1mm}
E_x^>(x',0)\, H^{(1)}_0(k_0\sqrt{(x \hspace*{-0.5mm} - \hspace*{-0.5mm} x')^2 +
\hspace*{-0.5mm} z^2})\,dx' .
\end{equation}
\noindent It generalizes the Huygens principle. The $H_0^{(1)}$ function describes a point irradiation source placed at $z' = 0$, and the field $E_x^>(x',0)$, which is nonzero for $|x'| < l$, serves as a density of the oscillating magnetic moment. The effective irradiation source replaces indeed the real sources -- the surface currents at the air-metal boundaries.

The case "In" can be treated similarly. In the subwavelength case, which is of prime interest, the propagating mode incident from the inside is fundamental. Repeating the calculations, we come to the symmetry relations for the amplitudes of the excited waves/modes: $a_k^{in} = - a_k$, $b_0^{in} = 1 - b_0$, and $b_{\nu}^{in} = - b_{\nu}$ for $\nu \neq 0$, where the amplitudes $a_k$ and $b_{\nu}$ are taken for the normal incidence. Thus, the case "In" is {\it reducible} to the case "Out".

Computation of the coefficients $T_{\nu \nu'}$ and determination of the amplitudes $b_{\nu}$ via truncation of Eq.~(\ref{MasterObl}) present no special problems, see below. However, the fundamental limit $l \to 0$ can be treated analytically. We have here $T_{00} \to 0$, $f_{\nu, -k_{0x}} \to 2\delta_{\nu 0}$, and, correspondingly, $b_0 \equiv \langle H^>(x,0) \rangle \to 2$ and $b_{\nu} \to 0$ for $\nu \neq 0$ for any $\theta_0$.

The dependence of the amplitude $b_{\nu}$ on the angle $\theta_0$ and the ratio $r = 4l/\lambda$ exhibits {\it important features}. The cases of small and large angle of incidence are different, see Fig.~2. For $\theta_0 \lesssim 10^{\circ}$, the situation is close to that for $\theta_0 = 0$: The fundamental $0$-mode dominates everywhere, the function $|b_0|(r)$ drops from $2$ to $\simeq 1$ and then remains almost constant, while $|b_{1,2, \ldots}|(r) \ll 1$.
\begin{figure}[h]
\centering
\includegraphics[width=8.4cm,height=3cm]{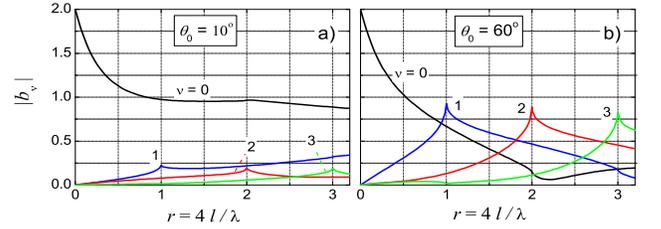}
\caption{The amplitudes $|b_{0,1,2,3}|$ versus $4l/\lambda$ for $\theta_0 = 10^{\circ}$ (a) and $60^{\circ}$ (b); the truncation number $\nu_{\rm max} = 20$.}\label{Fig2}
\end{figure}
For $\theta_0 \gtrsim 30^{\circ}$, the $0$-mode dominates only for $r \lesssim 0.5$. Otherwise, there is a strong mode competition. The opening of new propagating modes at $r = 1,2, \ldots$ is linked to sharp peaks of $|b_{1,2,\ldots}|(r)$; the odd and even peaks are comparable with each other tending to $1$ for $\theta_0 \to 90^{\circ}$. Mutual influence only of the modes of the same parity is evident. Only a few nearest modes dominate for each particular value of $r$, i.e. a {\it selective mode excitation} takes place.

Consider now the interface characteristics. The transmission properties can be described by the efficiency
\begin{equation}\label{EtatGen}
\eta_t(r,\theta_0) = \sum_{\nu} \; c_{\nu}\,|b_{\nu}|^2\; \mbox{Re}\hspace*{0.3mm} \beta_{\nu} \, / \, k_0\cos \theta_0 \;,
\end{equation}
\noindent which is the ratio of the energy flux through the slit to the flux incident onto the slit. The transmission efficiency is the sum of the partial contributions from all propagating modes. For $r \leq 1$ we have $\eta_t(r,\theta_0) = |b_0|^2/\cos \theta_0$; this gives $\eta_t(0,\theta_0) = 4/\cos\theta_0$ for $r \to 0$. Similarly, we introduce the differential diffraction efficiency
\begin{equation}\label{etad}
\eta_d(r,\theta,\theta_0) = \pi k_0 \cos^2\theta\;|a_k|^2 / \hspace*{1mm} l\cos \theta_0 \;,
\end{equation}
\noindent where $\theta = \arcsin (k/k_0)$ is the diffraction angle. Owing to the symmetry properties, we have $\eta_d^{in}(r,\theta) = \eta_d(r,\theta,0)$. Integrating $\eta_d (r,\theta,\theta_0)$ over $\theta$, we obtain the total diffraction efficiency $\eta_d^{\Sigma} (r,\theta_0)$. The products $2l\cos \theta_0\,\eta_{t,d}$ give the cross-sections of the corresponding transformation processes, and the difference $R_{in}(r) = 1 - \eta_d^{\Sigma}(r,0)$ gives the internal-reflection coefficient for the problem "In". In the limit $r \to 0$ we have $\eta_d^{\Sigma} (0,\theta_0) = 0$ and $R_{in}(0) = 1$.

Owing to the non-dissipative character of the transformation processes, the amplitudes $b_{\nu}$ obey the energy conservation law, which is similar to the "optical theorem" of the scattering theory. It looks especially simple for the normal incidence: $\eta_t + \eta_d^{\Sigma} = 2\mbox{Re} \hspace*{0.4mm} b_0$, which allows to express the phase of $b_0$ by $\eta_t$ and $\eta_d^{\Sigma}$ for $r < 1$.

Concerning the truncation procedure, it is sufficient to take into account $2 - 3$ evanescent modes in addition to the propagation modes in order to achieve a sub-percent accuracy for $\eta_{t,d}$ and to fulfill nicely the energy conservation law. This means that $\nu_{\rm max} \approx [ 4l/\lambda ] + 3$.

Fig.~3 shows the behavior of $\eta_t$ and $\eta_d^{\Sigma}$ when changing $r = 4l/\lambda$ and $\theta_0$. It should be considered in co-junction with Fig.~2. With $\eta_t(0) = 4/\cos \theta_0$, the function $\eta_t(r)$ is decreasing up to $r = 1$. Further increase of $r$ results in bursts at the openings of new propagating modes. For $\theta_0 = 0$, small bursts occur only at $r \simeq 2,4, \ldots$ With increasing $\theta_0$, they appear also at $r \simeq 1, 3, \ldots$ and become all highly pronounced and asymmetric. The asymmetry is due to a sharp square-root-law growth of the propagating constants $\beta_{\nu}(r)$ for $r > \nu$.
\begin{figure}[h]
\centering
\includegraphics[width=8.4cm,height=3cm]{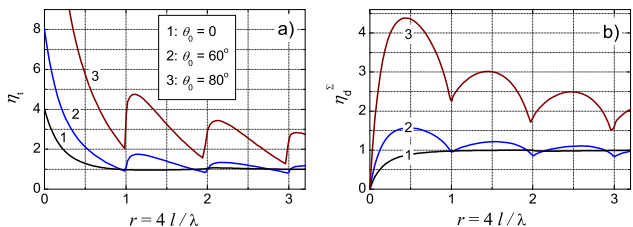}
\caption{ The transmission efficiency (a) and the total diffraction efficiency (b) versus $4l/\lambda$ for $\theta_0 = 0$, $60^{\circ}$, and $80^{\circ}$.}\label{Fig3}
\end{figure}
The function $\eta_d^{\Sigma}(r)$ grows initially with a $\theta_0$-dependent slope. The further scenario depends on the angle $\theta_0$. For $\theta_0 \ll 1$, one sees only small bursts. Correspondingly, with increasing $r$, the reflection coefficient for the problem "In", $R_{in}(r) = 1 - \eta_d^{\Sigma} (r,0)$, decreases from $1$ to almost $0$. For large $\theta_0$, the first maximum of $\eta_d^{\Sigma}(r,\theta_0)$, situated deeply in the subwavelength range, considerably exceeds~$1$, and the subsequent oscillations are highly pronounced and almost symmetric. Sharp minima of $\eta_d^{\Sigma}(r)$ occur at zeros of $\beta_{\nu}(r)$, i.e., at the sharp maxima of $|b_{\nu}|(r)$ in Fig.~2b.

\begin{figure}[h]
\centering
\includegraphics[width=8.4cm,height=3cm]{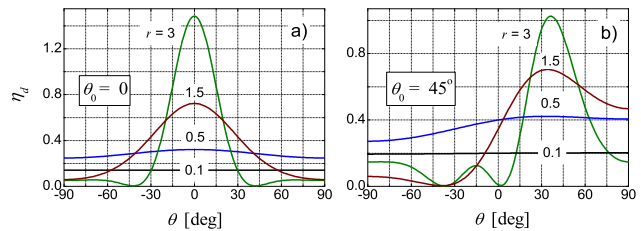}
\caption{The differential diffraction efficiency for $\theta_0 = 0$ (a) and $45^{\circ}$ (b).
The curves refer to $r = 0.1,\, 0.5,\, 1.5$, and $3$.}\label{Fig4}
\end{figure}

Fig.~4 shows what happens to the angular dependence $\eta_d(\theta)$ when increasing $r$ and $\theta_0$. For $\theta_0 = 0$, the quasi-isotropic distribution $\eta_d(\theta) \simeq const$, which occurs for $r \lesssim 1$, transforms gradually into a central peak growing and narrowing for $r \gg 1$. After the opening of the second propagating mode ($r > 1$) this peak acquires oscillating tails. For $\theta_0 > 0$, the main difference in the behavior is in a progressive shift of the peak to the right with increasing $r$. For $r \gg 1$ the value of $\theta_{\rm peak}$ approaches $\theta_0$ and the diffracted waves concentrate around the reflected one.

As seen from Fig.~4, the differential diffraction efficiency remains non-zero for $\theta \to \pm 90^{\circ}$, i.e. for ultimately large diffraction angles. This ``grazing diffraction" grows remarkably and becomes strongly asymmetric with increasing $\theta_0$: $\eta_d(90^{\circ}) \gg \eta_d(-90^{\circ})$. Depending on $r$, the maxima of $\eta_d(\pm 90^{\circ})$ correlate to those of $\eta_d^{\Sigma}(r)$. The "grazing" diffraction is closely related to the excitation of the surface plasmons in real metals, see also below.

The interface characteristics considered are sufficient to describe the transmission/diffraction properties of a subwavelength slit in an opaque film of a thickness $d$~\cite{Garcia02,We3}. Internal reflections of the fundamental propagating mode from the opposite interfaces lead to a sequence of the Fabry-Perot resonances; their positions are controlled by the $k_0d$-product. At a resonance, the total-transmission efficiency is $\eta_t/(1 - R_{in})$ and the intensity amplification factor inside the slit is $\eta_t/(1 - R_{in})^2$. These quantities grow rapidly with decreasing~$r$. At $\theta_0 = 0$, the resonant transmission cross-section $2l\,\eta_t /(1 - R_{in})$ tends to $\lambda/2$ for $r \to 0$.

In addition to $\eta_{t,d}$, the spatial profiles $H^<(x,0)$ and $E^<_{x,z}(x,0)$ show intimate features of the subwavelength behavior. Fig.~5 shows these profiles for the normal incidence, a few representative values of~$k_0l$, and $\nu_{\rm max} = 200$. So high values of the truncation number are needed solely to resolve the corner singularities at $|x| = l$.

The profile $H^<(x,0)$, see Figs.~5a and 5b, consists of two distinct sections (inside and outside the slit) separated by the inflection points at $|x| = l$. In the subwavelength range, the function $H^<(x,0)$ is structureless inside the slit and tending to the limiting value $H(\infty,0) = 2$ for $k_0l \to 0$, which corresponds to the slitless case. The same limiting value takes place for  $\theta_0 \neq 0$. This is why we have $b_0 \to 2$ and $\eta_t \to 4/\cos \theta_0$ for $k_0l \to 0$. For $k_0l \sim 1$, we have already $H^<(x,0) \simeq 1$, which corresponds to a weakly perturbed entering into the slit and to $\eta_t \approx 1$, see Fig.~3a. For $k_0l = 5$, the corner value $H^<(l,0)$ is close to $4/3$, which corresponds to the Sommerfeld solution for a single metal wedge~\cite{Sommerfeld}. Larger values of $H^<(l,0)$ for $k_0l \lesssim 1$ are due to the mutual influence of the wedges.

Fig.~5c shows the near-field behavior of $|E^<_{x}(x,0)|$ and $|E^<_{z}(x,0)|$. The first function, serving as the radiation source in Eq.~(\ref{deltaH}), is even in $x$ and zero for $|x| > l$. The second function, which is odd in $x$, is not restricted to the slit area. At the inflection points, they both tend to infinity as $|x - l|^{-1/3}$, which corresponds to the $90^{\circ}$ corner singularities of the ideal metal~\cite{Landau,Jackson}. The larger $\nu_{max}$, the clearer is the singular behavior.

\begin{figure}[h]
\centering
\includegraphics[width=8.5cm,height=5.1cm]{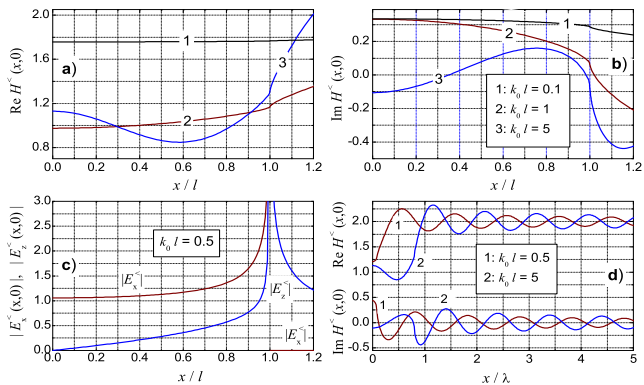}
\caption{Dependences $H^<(x,0)$ and $E^<_{x,z}(x,0)$ in the near and far fields. Curves 1, 2, and 3 in a), b) are plotted for $k_0l = 0.1$, $1$, and $5$. Curves 1 and 2 in c) correspond to $k_0l = 0.5$. The far-field profiles $1$ and $2$ in d) refer to $k_0l = 0.5$ and $5$. }\label{Fig5}
\end{figure}

Fig.~5d illustrates the far-field behavior. Outside the slit, where the characteristic spatial scale is $\lambda$, we see a quickly establishing radiation pattern superimposed on the background value $H^<(\infty,0) = 2$. The intensity $|H^<(x,0) - 2|^2$ decreases as $1/|x|$ without oscillations for $x - l \gtrsim \lambda$, as expected for the 2D case. Furthermore, it grows and then oscillates with increasing $k_0l$.

Several issues are worthy of further discussion. \\
-- Our results show that the growth of the transmission efficiency $\eta_t(r,\theta_0)$ for $r \equiv 4l/\lambda \to 0$ and $\theta_0 \to \pi/2$ is linked to the general feature of the near-field behavior: The average over the slit $\langle H(x,0) \rangle$ tends to $2$ (the slitless limit) for $r \to 0$ leading to the amplitude of the propagating mode $b_0 = 2$ for any $\theta_0$. \\
-- The presence of noticeable grazing diffraction, see Fig.~5d, elucidates the mechanism of surface-plasmon generation in real metals. The radiation mechanism, pertaining near the slit, is almost the same for the ideal metal and for real metals with large negative values of the optical permittivity~$\varepsilon_m$. The presence of the localized surface mode in the latter allows to catch the waves diffracted at large angles. The surface-plasmon excitation efficiency can be evaluated as $\eta_{\rm sp} \simeq f\,\eta_d(\pi/2)$, where $f = 2/\sqrt{|\varepsilon_m|}$ is the
numerical aperture. This estimate is in good agreement with~\cite{LalannePRL05}.  \\
-- To a big extent, the ideal-metal model is applicable to real metals with $|\varepsilon_m| \gg 1$. The single-slit characteristics experience only minor changes for $k_0l > |\varepsilon_m|^{-1/2}$, when the slit width exceeds the skin depth. In many cases $|\varepsilon_m| = 10^1 - 10^2$, and this inequality is not restrictive. For $l/\lambda \to 0$, the single-slit characteristics, especially $\eta_t$, can experience substantial changes. \\
-- The corner singularities are clearly seen in the near-field for $\nu_{\rm max} \gtrsim 10^2$. However, they are uncoupled from the diffraction/transmission properties. The latter are linked to several lowest modes, i.e., to rough features of the near-field behavior. A small-radius edge rounding is thus not expected to produce a strong effect on $\eta_{t,d}$.

In conclusion, a full-scale physical picture of the transmission, diffraction, and near-field properties of a single slit in ideal metal is presented. In the sub-to-near-subwavelength range, the transformation efficiencies show sharp dependences on $l/\lambda$ and $\theta_0$ which are closely linked to the near-field behavior. For skin-thick films, the sharp subwavelength behavior leads to a strong Fabry-Perot enhancement of the total transmittance, and of light inside the slit. Oblique incidence strongly facilitates the excitation of the non-zero eigenmodes leading to mode competition. Only few selectively excited modes are strongly involved in the near-subwavelength transmission/diffraction phenomena. The results obtained serve as a reference point in nanooptics of metals.

\vspace*{1mm} \noindent {\bf Acknowledgement:} Financial support from the Programs of Presidium RAN 21.2 and BPS of RAN "Physics of new materials and structures" is acknowledged.

\end{document}